\documentclass{article}
\usepackage{amssymb}
\usepackage{amsmath}
\def\R{{\mathbb R}}

\usepackage{graphicx}

\begin{document}

\title{Analysis of a model of the Calvin cycle with diffusion of ATP}

\author{Burcu G\"urb\"uz\\
and\\
Alan D. Rendall\\
Institut f\"ur Mathematik\\
Johannes Gutenberg-Universit\"at\\
Staudingerweg 9\\
D-55099 Mainz\\
Germany}

\date{}

\maketitle

\begin{abstract}
The dynamics of a mathematical model of the Calvin cycle, which is part of
photosynthesis, is analysed. Since diffusion of ATP is included in the model
a system of reaction-diffusion equations is obtained. It is proved that
for a suitable choice of parameters there exist spatially inhomogeneous
positive steady states, in fact infinitely many of them. It is also shown that
all positive steady states, homogeneous and inhomogeneous, are nonlinearly
unstable. The only smooth steady state which could be stable is a trivial
one, where all concentrations except that of ATP are zero. It is found that
in the spatially homogeneous case there are steady states with the property
that the linearization about that state has eigenvalues which are not real,
indicating the presence of oscillations. Numerical simulations exhibit
solutions for which the concentrations are not monotone functions of time.
\end{abstract}  

\section{Introduction}\label{intro}

Photosynthesis consists of two main parts, the light reactions and the dark
reactions. (For the basic facts about photosynthesis see \cite{alberts08}.)
The light reactions are the part in which light energy is captured
and used to produce energetic small molecules such as ATP, with oxygen as a
by-product. In the dark reactions this stored energy is used to produce sugars
starting from carbon dioxide. The chemical reaction network of the dark
reactions is often known as the Calvin cycle since its structure was
elucidated by Melvin Calvin and his collaborators.

There is a wide variety of mathematical models for the Calvin cycle in the
literature, most of which are systems of ordinary differential equations.
For reviews see \cite{jablonsky11}, \cite{arnold14} and
\cite{rendall17}. One exception to this is a model
introduced in \cite{grimbs11} which is a system of reaction-diffusion
equations. In fact in that model the only substance whose diffusion is taken
into account is ATP. The motivation for this model is as follows. There
were indications that the Calvin cycle might admit more than one 
steady state. If this were the case it might be of practical interest, since
it opens up the prospect of modifying the circumstances of photosynthesis so
as to move the process to a different steady state. In the best case this
steady state would exhibit a higher yield of sugars than the usual
one and could be used to increase food production. In a simple ODE model
of the Calvin cycle introduced in \cite{grimbs11} there is only one steady
state and the authors introduced a hypothesis as to how this could be changed
by a modification of the model. The idea was to take into account the
diffusion of ATP. Steady states of the original system then correspond to
spatially homogeneous steady states of the modified one. Even if there
only exists one homogeneous steady state there might also exist 
inhomogeneous steady states. In experiments which are only able to
determine spatial averages of the concentrations of the substances involved
the fact that a steady state is inhomogeneous would not be visible.

With this background we study the model of \cite{grimbs11} with diffusion in
what follows. The primary aim is to investigate the existence of inhomogeneous
steady states. The biological interest of steady states is dependent on
their stability and so we also look at the stability of homogeneous and
inhomogeneous steady states. In addition we study the existence of
oscillations in this model. The paper is organized
as follows. In Section \ref{model} the model is introduced and some of
its general properties are discussed. The existence of inhomogeneous steady
states is proved in Theorem 2 of Section \ref{existence}. Section
\ref{global} contains a proof of global in time existence for dynamical
solutions and some statements about global bounds for these solutions. In
particular Theorem 3 describes the long-time behaviour of solutions for a
restricted set of parameters. In Section \ref{stability}
it is shown that all positive steady states, homogeneous and inhomogeneous, are
(nonlinearly) unstable as solutions of the model with diffusion. Oscillations
in the model are studied in Section \ref{oscillations}. This is done for
spatially homogeneous solutions of the system of reaction-diffusion equations
and for solutions of a related ODE model where the concentration of ATP is
taken to be fixed. Cases are found where the linearization about a steady
state has eigenvalues which are not real, indicating the occurrence of
oscillations. Numerical simulations are carried out which exhibit solutions
where the concentrations of the substances involved depend on time in a
non-monotone manner.

\section{The model}\label{model}

The model of central interest in this paper is a mathematical description of
the Calvin cycle of photosynthesis including diffusion. It is a system of
reaction-diffusion equations which was introduced in \cite{grimbs11} (system
(13) of that paper). The
equations are
\begin{eqnarray}
  &&\frac{\partial x_{\rm RuBP}}{\partial t}=-k_1x_{\rm RuBP}+k_5x_{\rm Ru5P}
  x_{\rm ATP},\label{mad1}\\
  &&\frac{\partial x_{\rm PGA}}{\partial t}=2k_1x_{\rm RuBP}-k_2x_{\rm PGA}x_{\rm ATP}
  -k_6x_{\rm PGA},\label{mad2}\\
  &&\frac{\partial x_{\rm DPGA}}{\partial t}=-k_3x_{\rm DPGA}
     +k_2x_{\rm PGA}x_{\rm ATP},\label{mad3}\\
  &&\frac{\partial x_{\rm GAP}}{\partial t}=k_3x_{\rm DPGA}-k_7x_{\rm GAP}
  -5k_4x_{\rm GAP}^5,\label{mad4}\\
  &&\frac{\partial x_{\rm Ru5P}}{\partial t}=3k_4x_{\rm GAP}^5
     -k_5x_{\rm Ru5P}x_{\rm ATP},
  \label{mad5}\\
  &&\frac{\partial x_{\rm ATP}}{\partial t}=-k_2x_{\rm PGA}x_{\rm ATP}
     -k_5x_{\rm Ru5P}x_{\rm ATP}+k_8(c-x_{\rm ATP})
     +D_{\rm ATP}\frac{\partial^2x_{\rm ATP}}{\partial R^2}.\label{mad6}
\end{eqnarray}
The unknowns here are the concentrations of the substances RuBP (ribulose
1,5-bisphosphate), PGA (phosphoglycerate), DPGA (1,3-diphosphoglycerate),
GAP (glyceraldehyde 3-phosphate), Ru5P (ribulose 5-phosphate) and
ATP (adenosine triphosphate), with $x_X$ denoting the concentration of the
substance X.
When it seems helpful we will replace the notation $x_{\rm RuBP}$, $x_{\rm PGA}$,
$x_{\rm DPGA}$, $x_{\rm GAP}$, $x_{\rm Ru5P}$, $x_{\rm ATP}$ by the more convenient
but less informative notation $u_i$, $1\le i\le 6$, or $u$ for the
vector-valued function with components $u_i$.
The concentrations are functions of time $t$ and one spatial coordinate
$R$. The rate constants $k_i$ are positive real numbers as are $D_{\rm ATP}$,
the diffusion coefficient of ATP, and $c$, the total amount of adenosine
phosphates.

From the point of view of the biological applications it would be natural
to consider a region in three-dimensional space. Here we reduce to one spatial
dimension so as to have the simplest situation in which the effects of
diffusion may be seen. The biologically natural boundary conditions,
corresponding to the case where the substances involved cannot pass
through the boundary, are Neumann conditions, where the derivative with
respect to $R$ vanishes. Considering a general situation with these boundary
conditions leads to some mathematical technicalities. In particular, the
initial data have to satisfy some restrictions at the corner points where the
initial hypersurface meets the spatial boundary. These conditions are not likely
to lead to any insight into the problem we are considering. An alternative
would be to simplify by replacing the boundary consitions by the assumption
that the unknowns are periodic functions of $R$ with period $\eta$. In that
case the unknowns are defined at points $(R,t)$ of a subset of
$\R\times [0,\infty)$ and satisfy the condition $u(R+\eta,t)=u(R,t)$ for all
$R$ and $t$. Initial data are specified for $t=0$ and are assumed to be
periodic. In what follows we will consider the case of Neumann boundary
conditions except when the contrary is stated and will supplement that by
a few remarks on the case of periodic boundary conditions.

Equations (\ref{mad1})-(\ref{mad6}) are referred to in what follows, as in
\cite{rendall14}, as the MAd system (mass action with diffusion). Dropping the
last equation and setting the concentration of ATP to a constant value in the
other equations leads to a system of five ODE which we refer to as the MA
system and which was the starting point of the considerations in
\cite{grimbs11} (system (5) in that paper). In that context the constant
factor of the concentration of ATP can be absorbed into the definition of the
rate constants $k_2$ and $k_5$. Due to their
interpretation as concentrations the unknowns are assumed to be non-negative.
In addition it is assumed that $x_{\rm ATP}\le c$. This is because $c-x_{\rm ATP}$
represents the concentration of ADP. The simplified system where the last
summand in (\ref{mad6}) is omitted is obtained if diffusion is neglected
($D_{\rm ATP}=0$) or if the concentrations, in particular that of ATP, are
spatially homogeneous, i.e. only depend on time. The latter system is called,
as in \cite{rendall14}, the MAdh system (mass action with diffusion,
homogeneous).

A basic existence theorem for the MAd system will now be stated. By a classical
solution of the initial boundary value problem we mean a continuous
function $u$ for which $\frac{\partial u}{\partial t}$ and
$\frac{\partial^2 u_6}{\partial R^2}$ exist
and are continuous and the equations (\ref{mad1})-(\ref{mad6}) and the
boundary conditions hold pointwise. The formulation of the theorem uses
certain spaces of functions which have different degrees of H\"older
regularity with respect to the space and time variables. Their precise
definitions will not be given here and we refer to \cite{rothe84} and its
references for more details. The H\"older coefficient $\alpha$ satisfies
$0<\alpha<1$.

\noindent
{\bf Theorem 1} Let $u_0:[R_1,R_2]\to\R^6$, be a non-negative function on an
interval with $R_1<R_2$, where the components $u_{0,i}$ are of class $C^{\alpha}$
for $1\le i\le 5$ and $u_{0,6}$ is of class $C^{2+\alpha}$. Suppose that
$u_{0,6}\le c$ and that $u_{0,6}$
satisfies Neumann boundary conditions. Then there exists a $T>0$ and a
unique function $u(R,t)$ on $[R_1,R_2]\times [0,T)$ where $u_i$ is of class
$C^{\alpha,1+\frac{\alpha}{2}}$ for $1\le i\le 5$ and $u_6$ is of class
$C^{2+\alpha,1+\frac{\alpha}{2}}$, $u(R,0)=u_0(R)$ for all $R$ and $u$ is a
classical solution of (\ref{mad1})-(\ref{mad6}), with $u_6$ satisfying Neumann
boundary conditions. If for any solution of this type with given initial
datum and finite $T$ the quantity $\|u(t)\|_{L^\infty}$ is bounded then there
exists a classical solution with the given initial datum on
$[R_1,R_2]\times [0,\infty)$. The $u_i$ are non-negative and $u_6\le c$.

\noindent
{\bf Proof} All the statements in this theorem except those contained in the
last sentence follow from \cite{rothe84}, Part II, Theorem 1. To prove those
consider first the case that $u_{0,i}>0$  for $1\le i\le 6$ and $u_{0,6}<c$.
Then $u_i(t)>0$ and $u_6(t)<c$ for $t$ sufficiently small. Let $t_1$ be the
supremum of the times such that these inequalities hold for $t\in [0,t_1)$.
Assume that $t_1<T$ in order to obtain a contradiction. Then either there exists
some $i<6$ and some $R$ such that $u_i(t_1,R)=0$ or there exists some $R$ such
that $u_6(t_1,R)=0$ or $u_6(t_1,R)=c$.  In the case $i<6$ the function $u_i$
solves an equation of the form $\frac{du_i}{dt}=f(t)u_i+g(t)$ for continuous
functions
$f$ and $g$ with $g\ge 0$. On the interval $[0,t_1)$ the function $u_i$ is
positive and we obtain an inequality of the form $\frac{d}{dt}(\log u_i)\ge f$.
Hence $u_i$ remains bounded away from zero as $t\to t_1$, a contradiction. So
in fact for $1\le i\le 5$ the function $u_i$ is positive for $t=t_1$. Next
consider the case that $u_6$ approaches zero as $t\to t_1$. It satisfies an
inequality of the form
\begin{equation}\label{maxineq1}
-\frac{\partial u_6}{\partial t}+D_{\rm ATP}\frac{\partial^2 u_6}{\partial R^2}
-f(t)u_6\le 0
\end{equation}
for a positive continuous function $f$. The maximum principle
(cf. \cite{protter84}, Chapter 3, Theorem 4) gives a contradiction. Finally,
consider the case where $u_6$ approaches $c$ as $t\to t_1$. We have the
inequality
\begin{equation}\label{maxineq2}
-\frac{\partial(u_6-c)}{\partial t}
+D_{\rm ATP}\frac{\partial^2(u_6-c)}{\partial R^2}-f(t)(u_6-c)\ge 0
\end{equation}
and once again the maximum principle
gives a contradiction. This completes the proof in the case that it is
assumed that the initial data satisfy the strict inequalities. If the data
only satisfy the non-strict inequalities then they can be uniformly approximated
by a sequence of data satisfying the strict inequalities. It follows in a
straightforward way from the proof of the existence theorem that the
corresponding sequence of solutions, which satisfy the strict inequalities,
converge to the solution corresponding to the original initial data. Hence
the latter solution also satisfies the non-strict inequalities. $\blacksquare$

\noindent
{\it Remark} An analogue of this result in the case of periodic boundary 
conditions which requires less sophisticated regularity assumptions on the 
initial data follows from results in Section 14A of \cite{smoller94}.


\section{Existence of steady states}\label{existence}

We begin by considering homogeneous steady states of the MAd system, i.e. steady
states of the MAdh system. The equations for these steady states are a system
of six algebraic equations. We are also interested in the stability of the
steady states. The results on these questions which are available in the
literature will now be summarized. In \cite{rendall14} it was shown that for
arbitrary values of the parameters there are at most two positive steady
states of the MAdh system. When $c<5k_6/k_2$ there are none. When $c$ is
sufficiently large for fixed values of the rate constants there are two.
There is one value $c^*$ of $c$ where the two steady states coalesce. In that
case there is precisely one steady state and it is degenerate. It was shown in
\cite{disselnkoetter17} that there are parameter values for which there is one
stable and one unstable steady state. It was left open whether this is
true for all parameter values for which there are two steady states. 

A natural way to try to understand more about the solutions of the equations
for steady states is to eliminate as many unknowns as possible. Equations
(14) in \cite{grimbs11} express all the other concentrations in terms of that
of RuBP. We will now derive these equations, together with some others. Note
that adding (\ref{mad1}) and (\ref{mad5}) gives $3k_4x_{\rm GAP}^5=k_1x_{\rm RuBP}$
and hence 
\begin{equation}\label{gaprubp}
x_{\rm GAP}=\left(\frac{k_1x_{\rm RuBP}}{3k_4}\right)^{\frac15},
\end{equation}
which is the third equation of the system (14) in \cite{grimbs11}.
Combining (\ref{mad3}), (\ref{mad4}) and (\ref{mad5}) leads to the relation
\begin{equation}
k_2x_{\rm PGA}x_{\rm ATP}+k_5x_{\rm Ru5P}x_{\rm ATP}=k_7x_{\rm GAP}+8k_4x_{\rm GAP}^5.
\end{equation}  
This can be used together with (\ref{mad6}) to express the concentration of
ATP in terms of that of RuBP with the result that
\begin{equation}\label{atp}
  x_{\rm ATP}=c-\frac{8k_1x_{\rm RuBP}}{3k_8}-\frac{k_7}{k_8}
  \left(\frac{k_1x_{\rm RuBP}}{3k_4}\right)^{\frac15}.
\end{equation} 
This is the fifth equation of the system (14) in \cite{grimbs11}. Combining 
(\ref{mad5}) and (\ref{gaprubp}) and substituting (\ref{atp}) into the result
gives
\begin{equation}\label{ru5p}
  x_{\rm Ru5P}=\frac{3k_1k_8x_{\rm RuBP}}
  {k_5(3k_8c-8k_1x_{\rm RuBP}-3k_7
  \left(\frac{k_1x_{\rm RuBP}}{3k_4}\right)^{\frac15})},
\end{equation}
which is the fourth equation of the system (14) in \cite{grimbs11}. Using
(\ref{mad4}) and (\ref{gaprubp}) gives
\begin{equation}\label{dpga}
  x_{\rm DPGA}=\frac{5k_1x_{\rm RuBP}
    +3k_7\left(\frac{k_1x_{\rm RuBP}}{3k_4}\right)^{\frac15}}{3k_3}
\end{equation}
which is the second equation of the system (14) in \cite{grimbs11}. Combining
(\ref{mad3}) and (\ref{atp}) gives
\begin{equation}\label{pga}
x_{\rm PGA}=\frac{5k_1x_{\rm RuBP}
  +3k_7\left(\frac{k_1x_{\rm RuBP}}{3k_4}\right)^{\frac15}}
{k_2(3c-\frac{8k_1x_{\rm RuBP}}{k_8}-\frac{3k_7}{k_8}
  \left(\frac{k_1x_{\rm RuBP}}{3k_4}\right)^{\frac15})}
\end{equation}
which is the first equation of the system (14) in \cite{grimbs11}. Note that
equation (\ref{mad2}) has not been used in the derivation of these equations
and as a consequence they are independent of $k_6$.

In \cite{rendall14} all other concentrations were expressed in a different way
in terms of $x_{\rm RuBP}$. There is no contradiction here - both systems of
equations are valid. In fact (\ref{atp}) is the same as equation (6.16) in
\cite{rendall14} and (\ref{gaprubp}) is obviously equivalent to an equation in
\cite{rendall14}. The other relevant equations in \cite{rendall14} are
\begin{eqnarray}
  &&x_{\rm PGA}=\frac{2k_1x_{\rm RuBP}}{k_2x_{\rm ATP}+k_6},\label{ss1}\\
  &&x_{\rm DPGA}=\frac{k_7x_{\rm GAP}}{k_3}+\frac{5k_4x_{\rm GAP}^5}{k_3},
     \label{ss2}\\
  &&x_{\rm Ru5P}=\frac{3k_4x_{\rm GAP}^5}{k_5x_{\rm ATP}}.\label{ss4}
\end{eqnarray}
Here $x_{\rm GAP}$ and $x_{\rm ATP}$ have not been eliminated everywhere in favour
of $x_{\rm RuBP}$ but it is clear how to do so if desired.

It is easy to analyse steady states of the MAdh system on the boundary of the
positive orthant. Note first that at such a steady state the concentration of
ATP cannot vanish. It is also the case that there is a cyclic implication of
vanishing of the concentrations of RuBP, Ru5P, GAP, DPGA, PGA, Ru5P, RuBP. Hence
either all of these concentrations are non-zero or they are all zero. For a
steady state on the boundary the latter must hold and then $x_{\rm ATP}=c$. Thus
we see that there is a unique steady state on the boundary and that its
coordinates are $(0,0,0,0,0,c)$. The cyclic implication of vanishing also holds
at any $\omega$-limit point of a positive solution and so the only point which
can occur as a limit point of this kind is the steady state already mentioned.


In \cite{grimbs11} some explicit formulae are given which hold in any steady
state of the MA system. These are easily adapted to the case of the MAdh
system, giving
\begin{eqnarray}
  && x_{\rm RuBP}=k_1^{-1}
     \left[\frac{k_7^5}{3k_4\left(\frac{2k_2x_{\rm ATP}}{k_2x_{\rm ATP}+k_6}
   -\frac53\right)^5}\right]^{\frac14},\label{matrans1}\\
  &&x_{\rm Ru5P}=\frac{k_1}{k_5x_{\rm ATP}}x_{\rm RuBP},\label{matrans2}\\
  &&x_{\rm PGA}=\frac{2k_1}{k_2x_{\rm ATP}+k_6}x_{\rm RuBP},\label{matrans3}\\
  &&x_{\rm DPGA}=\frac{2k_1k_2x_{\rm ATP}}{k_3(k_2x_{\rm ATP}+k_6)}
     x_{\rm RuBP}.\label{matrans4}
\end{eqnarray}
Note that in the derivation of (\ref{matrans1}) it is assumed that
$x_{\rm RuBP}\ne 0$. Together with (\ref{gaprubp}) these equations show that at
steady state all other concentrations can be expressed in terms of that of ATP.

In the case of homogeneous steady states of the MAd system
substituting (\ref{matrans1}) into (\ref{atp}) gives
$x_{\rm ATP}=c-\frac{8k_4}{k_8}f(x_{\rm ATP})
-\frac{k_7}{k_8}[f(x_{\rm ATP})]^5$ where
\begin{equation}
  f(x_{\rm ATP})=\left[\frac{k_7}
    {3k_4\left(\frac{2k_2x_{\rm ATP}}{k_2x_{\rm ATP}+k_6}
   -\frac53\right)}\right]^{\frac54}.
\end{equation}
The function $f$ is only defined for $x_{\rm ATP}>\frac{5k_6}{k_2}$ and tends to
infinity as $x_{\rm ATP}\to x^*=\frac{5k_6}{k_2}$.  As
$x_{\rm ATP}\to\infty$ we have $f(x_{\rm ATP})\to
\left(\frac{k_7}{k_4}\right)^{\frac54}$. $f$ is strictly decreasing. Explicitly
\begin{equation}
  f'(x_{\rm ATP})=-\frac{15k_2k_4k_6}{2k_7}
\left[\frac{k_7}
    {3k_4\left(\frac{2k_2x_{\rm ATP}}{k_2x_{\rm ATP}+k_6}
   -\frac53\right)}\right]^{\frac94}\frac{1}{(k_2x_{\rm ATP}+k_6)^2}<0.  
\end{equation}
It follows that $f'(x_{\rm ATP})\to -\infty$ as $x_{\rm ATP}\to x^*$ and
$f'(x_{\rm ATP})\to 0$ as $x_{\rm ATP}\to\infty$.
The function $f$ is convex. Explicitly
\begin{eqnarray}
  &&f''(x_{\rm ATP})=\frac{405k_2^2k_4^2k_6^2}{4k_7^2}
   \left[\frac{k_7}
    {3k_4\left(\frac{2k_2x_{\rm ATP}}{k_2x_{\rm ATP}+k_6}
     -\frac53\right)}\right]^{\frac{13}4}\frac{1}{(k_2x_{\rm ATP}+k_6)^4}\nonumber\\
  &&+\frac{15k_2^2k_4k_6}{k_7}
   \left[\frac{k_7}
    {3k_4\left(\frac{2k_2x_{\rm ATP}}{k_2x_{\rm ATP}+k_6}
     -\frac53\right)}\right]^{\frac94}
     \frac{1}{(k_2x_{\rm ATP}+k_6)^3}\\
  &&=\frac{15k_2^2k_4k_6}{4k_7}
    \left[\frac{k_7}
    {3k_4\left(\frac{2k_2x_{\rm ATP}}{k_2x_{\rm ATP}+k_6}
     -\frac53\right)}\right]^{\frac94}\frac{1}{(k_2x_{\rm ATP}+k_6)^3}\\
  &&\times\left[\frac{4k_2x_{\rm ATP}+7k_6}
    {k_2x_{\rm ATP}-5k_6}\right]>0.
\end{eqnarray}
At a homogeneous steady state
\begin{equation}
c-x_{\rm ATP}=\frac{8k_4}{k_8}f(x_{\rm ATP})
+\frac{k_7}{k_8}[f(x_{\rm ATP})]^5.
\end{equation}
Any solution of this equation gives rise to a steady state. Due to the convexity
of the right hand side of this equation the number of solutions for fixed values
of the parameters is $0$, $1$ or $2$. Moreover, if $c$ is sufficiently large for
fixed values of the other parameters there are two solutions. This provides a
new proof of the result of \cite{rendall14} mentioned above.

Consider now a steady state of the MAd system which need not be spatially
homogeneous. It satisfies the equation
\begin{equation}
D_{\rm ATP}x_{\rm ATP}''+k_8c-k_8x_{\rm ATP}-8k_4f(x_{\rm ATP})-k_7[f(x_{\rm ATP})]^5=0.
\end{equation} 
where the prime denotes $\frac{d}{dR}$. Define
\begin{equation}
F(x)=k_8c-k_8x-8k_4f(x)-k_7[f(x)]^5.
\end{equation}
Let $V$ be a primitive of $F$, which we call the potential. We can deduce
qualitative
properties of $V$ from those of $F$. Note that the function $f$ diverges like
$(x-x^*)^{-5/4}$ as $x\to x^*$. $F$ diverges like $(x-x^*)^{-25/4}$ in
this limit. The singularity is not integrable and hence the potential $V$
diverges as $x\to x^*$, tending to $+\infty$. For $x\to\infty$ the
potential behaves asymptotically like $-k_8x^2$. When $c<c^*$ the function
$F$ has no zeroes and the potential is monotone decreasing. Steady states
correspond to horizontal line segments whose endpoints lie on the graph
of $V$ (cf. \cite{fife79}). When $V$ is monotone a line segment of this type
must be infinite
towards the right and so the corresponding steady state is unbounded. Since
in what follows we are only interested in bounded steady states we do not
consider this case further. Consider now the case $c>c^*$.

\noindent
{\bf Lemma 1} If $c>c^*$ then there are real numbers $x_1<x_2$ such that
$V$ has a non-degenerate local minimum at $x_1$ and a non-degenerate local
maximum at $x_2$. It has no other local extrema.

\noindent
{\bf Proof} When $c>c^*$ the derivative of the potential has two zeroes
$x_1<x_2$. It follows from the convexity of $f$ that $f'(x_2)>f'(x_1)$ and
hence that $F'(x_2)<F'(x_1)$. Since $F(x)$ is negative for $x$ close to
$x^*$ it follows that $F'(x_1)\ge 0$ and since $F(x)$ is negative for
$x$ large it follows that $F'(x_2)\le 0$. If $F'(x_1)$ were zero then
$F'(x)$ would be negative for all $x>x_1$. But his would imply that
$F(x_2)<0$, a contradiction. Thus in fact $F'(x_1)>0$. A similar argument
shows that $F'(x_2)<0$. Hence $V$ has a non-degenerate local minimum for
$x=x_1$ and a non-degenerate local maximum for $x=x_2$. These are the
only points where $F$ is zero and the statement about local extrema follows.
$\blacksquare$

Let $E_-=V(x_1)$ and $E_+=V(x_2)$. For any $E$ with $E_-<E<E_+$ there is a line
segment $V=E$ whose endpoints lie on the graph of $V$ and whose interior is 
above that graph. It defines a bounded inhomogeneous steady state which
satisfies $\frac12 (x_{\rm ATP}')^2+V(x_{\rm ATP})=E$. This can be interpreted as
a steady state with Neumann boundary conditions
on an interval $[R_1,R_2]$ of suitable length. For as an
endpoint of the line segment is approached $V(x_{\rm ATP})\to E$ and hence
$x_{\rm ATP}'\to 0$. If the other parameters are held fixed we can think of
$x_1$ and $x_2$ as functions of $c$. As $c\to\infty$ they behave in such a way
that $x_1(c)\to x^*$ and $x_2(c)\to\infty$. Hence $f(x_1(c))\to\infty$ and
$f(x_2(c))\to \left(\frac{k_7}{k_4}\right)^{\frac54}$. As $c$ increases the
function $F$ increases. Thus with a suitable choice of primitive $V$ is also
an increasing function of $c$.
 
We now want to estimate the length of the interval where the steady state
solution is defined. Let the line segment defining the solution be $[x_-,x_+]$.
We have the identity
\begin{equation}
\eta=R_2-R_1=\int_{x_-}^{x_+}[2(E-V(x))]^{-\frac12}dx.
\end{equation}
In this formula $\eta$ is the minimal length of the interval on which a
solution can be defined. We could also
think of this as a solution on an interval of length $k\eta$ for any natural
number $k$. There exists a solution of this type for any $E$ with
$V(x_1)<E<V(x_2)$. The quantity $\eta$ is a continuous function of $E$ and it
tends to infinity as $E\to V(x_2)$ because the local maximum is non-degenerate.

\noindent
{\bf Theorem 2} Consider the system (\ref{mad1})-(\ref{mad6}) with all
parameters fixed and $c>c^*$, and an interval $[R_1,R_2]$. Then there
exist infinitely many distinct inhomogeneous positive steady states on the
given interval.

\noindent
{\bf Proof} That there exists a steady state follows directly from the
discussion above. For a given potential there are solutions defined by line
segments whose lengths take on all values between zero and the maximal value
$x_2-x_0$, where $V(x_0)=V(x_2)$ and $x_0<x_2$. The minimal lengths $\eta$ of the
intervals where the corresponding solutions are defined take on all values
between zero and infinity. There is a solution on an interval of length
$\eta/k$ for each natural number $k$. Piecing together solutions of this type
gives a solution on an interval of length $\eta$. The solution constructed
directly from the line segment is monotone. Thus the solutions for different
values of $k$ have different numbers of extrema and are therefore distinct.
$\blacksquare$

\noindent
{\it Remark} In an analogous way a statement can be obtained about steady states
which are periodic in $R$.

In \cite{marciniak13} it has been observed that for a system analogous to the
one considered in the present paper there exists another type of inhomogeneous
steady states. They are non-negative but not strictly positive. They are also
not smooth and only satisfy the equations in a weak sense. Presumably
there also exist solutions of (\ref{mad1})-(\ref{mad6}) of this type. They
would have the following structure. Let $S_i$ be a finite increasing sequence
of numbers in the interval $[R_1,R_2]$ with $1\le i\le k$, $S_1=R_1$ and
$S_k=R_2$. The function $u_6$ is everywhere $C^1$ and its restriction to each
interval $[S_i,S_{i+1}]$ is smooth. It satisfies Neumann boundary conditions at
$R_1$ and $R_2$. On some of the subintervals $u$ is positive. On others $u_i=0$
for $1\le i\le 5$ and $D_{\rm ATP}u_6''+k_8(c-u_6)=0$. This issue will not be
investigated further here since this type of solution does not seem relevant for
modelling the Calvin cycle.

\section{Global existence and boundedness}\label{global}

In \cite{rendall14} it was shown that all solutions of the MAdh system exist
globally in the future. With the help of Theorem 1 this can be extended to
the case of the MAd system in a straightforward way. Note first that we only
consider data for which $x_{\rm ATP}$ is bounded by $c$ and then the solution
satisfies the same bound. To obtain global existence it suffices to show that
the other concentrations remain bounded for any solution on a finite time
interval $[0,T)$. The maximum of the quantities $u_1$, $u_2$, $u_3$ and
$u_4+u_5$ satisfies a linear integral inequality. Thus the
concentrations cannot grow faster than linearly and, in particular, remain
bounded on bounded intervals.

In \cite{rendall14} it was shown that all solutions of the MAdh system
are bounded in the future. Diffusion may have a destabilizing effect and so
it is not clear whether solutions of the MAd system are globally bounded.
An example where problems of this type occur in a system where only one
substance diffuses is given in \cite{marciniak13}. In that example the
concentration of the diffusing species is bounded pointwise while the
species which do not diffuse
are only bounded in $L^1$. In the case of the MAd system adding equations
(\ref{mad1}) and (\ref{mad6}) and integrating in space gives
\begin{eqnarray}
&&  \frac{d}{dt}\left(\int (u_1+u_6) dR\right)=-k_1\int u_1dR-k_2\int u_2u_6 dR
   +k_8\int (c-u_6) dR\nonumber\\
  &&\le -m\left(\int (u_1+u_6) dR\right)+k_8c(R_2-R_1)
\end{eqnarray}
where $m=\min\{k_1,k_8\}$. It follows that the $L^1$ norm of
$x_{\rm RuBP}+x_{\rm ATP}$ can be bounded above by the maximum of its initial
value and the quantity $\frac{k_8c(R_2-R_1)}{m}$. Call this $\hat x_{\rm RuBP}$.
It follows from (\ref{mad2}) that the $L^1$ norm of $x_{\rm PGA}$ can be
bounded by the maximum of its initial value and
$\hat x_{\rm PGA}=\frac{2k_1\hat x_{\rm RuBP}}{k_6}$. Similarly the $L^1$ norms of
$x_{\rm DPGA}$ and $x_{\rm GAP}$ can be bounded by the maximum of their initial
values and
\begin{equation}
  \hat x_{\rm DPGA}=\frac{k_2c\hat x_{\rm PGA}}{k_3}\ \ \ {\rm and}\ \ \
  \hat x_{\rm GAP}=\frac{k_3\hat x_{\rm DPGA}}{k_7},
\end{equation}
respectively. It is not clear whether the $L^1$ norm of $x_{\rm Ru5P}$ also
remains bounded. Some partial information can be obtained in the following
way.
\begin{eqnarray}
  &&\frac{d}{dt}(\sum_{i=1}^6\alpha_iu_i)=(2\alpha_2-\alpha_1)k_1 u_1
+[(\alpha_3-\alpha_2-\alpha_6)k_2u_6-\alpha_2k_6]u_2\nonumber\\
  &&  +(\alpha_4-\alpha_3)k_3u_3-\alpha_4k_7u_4
     +(3\alpha_5-5\alpha_4)k_4u_4^5\nonumber\\
  &&  +(\alpha_1-\alpha_5)k_5u_5u_6-\alpha_6k_8u_6+\alpha_6k_8c
     +D_{\rm ATP}\frac{\partial^2u_6}{\partial R^2}
\end{eqnarray}  
Let $\alpha_1=3$, $\alpha_2=1$, $\alpha_3=5$, $\alpha_4=4$, $\alpha_5=4$,
$\alpha_6=5$. Then we get the inequality
\begin{equation}
  \frac{d}{dt}(\sum_{i=1}^6\alpha_i\int u_i)\le \int[-k_1u_1-k_6u_2
  -k_3u_3-4k_7u_4-k_5u_6u_5-5k_8u_6+5k_8c]
\end{equation}
It follows that if $\liminf_{(R,t)\in[R_1,R_2]\times [0,\infty)}x_{\rm ATP}$ is positive
then the $L^1$ norm of $x_{\rm Ru5P}$ is bounded. However none of the arguments
given up to now exclude the possibility of a solution where at late times the
concentration of ATP becomes small somewhere and the total amount of Ru5P
becomes arbitrarily large. Note that a solution of this type cannot be
spatially homogeneous. For it would have to have an $\omega$ limit point
with $x_{\rm ATP}=0$ and this has already been ruled out.

In the case where $ck_2\le 5k_6$ it is possible to prove  more about the
long-time behaviour as shown in the following theorem.

\noindent
{\bf Theorem 3} Suppose that $ck_2\le 5k_6$. Then all solutions of the MAd
system are bounded and $u_6$ is bounded below by a positive constant.
Moreover all concentrations $u_i$ with $1\le i\le 5$ converge to zero for
$t\to\infty$.

\noindent
{\bf Proof} Consider a region defined by the inequalities
$0\le L_1\le a$ and $0\le u_6\le c$ where 
$L_1=u_1+\frac12 u_2+\frac35 u_3+\frac35 u_4+u_5$.
The aim is to show that this is an invariant region in the sense of Section
14B of \cite{smoller94}. Note that $L_1$ is a Lyapunov function considered in
\cite{rendall14} and that under the evolution of the MAdh system we have
$\frac{dL_1}{dt}=-\frac12\left(k_6-\frac15 k_2u_6\right)u_2-\frac35 k_7u_4$.
It follows from Theorem 14.11 of \cite{smoller94} that this region is
invariant, i.e. that any solution which starts in a region of this type
remains in it. Since any solution lies in this region for some $a$ boundedness
follows. It is also the case that the region defined by the inequalities
$0\le L_1\le a$ and $b\le u_6\le c$ is invariant for suitable choices of $a$
and $b$. It suffices to assume that $0<b\le\frac{k_8c}{2(k_2+k_5)a+k_8}$. It
can be concluded that for this choice of parameters $u_6$ is bounded below by a
positive constant.

To show that all concentrations $u_i$ with $1\le i\le 5$ converge to zero for
$t\to\infty$ consider $u(t,R)$
for a fixed $R$. The function $L_1(u(t,R))$ is non-increasing. Let $t_n$ be a
sequence tending to infinity as $n\to\infty$. We want to show that
$u_i(t_n,R)\to 0$ for $1\le i\le 5$. The sequence $u_i(t_n,R)$ is bounded and
so by passing to a subsequence we can assume that it converges to some $u_i^*$.
$\dot u_i(t,R)$ is bounded. Suppose that $u^*_4$ is positive. Then there will
be a sequence
of times where $u_4\ge\frac12 u^*_4$ and a sequence of intervals of length
$\delta>0$ where $u_4\ge\frac14 u^*_4$, which we can assume to be disjoint.
Each one of these intervals causes $L_1$ to decrease by a fixed amount and
this would mean that $L_1$ would eventually become negative, a contradiction.
Hence $u^*_4=0$. By an analogous argument $u^*_2=0$. If $u^*_3$ were positive
then $u_4$ would become arbitrarily large, a contradiction. Hence $u^*_3=0$.
Using the fact that $x_{\rm ATP}$ is bounded below by a positive constant 
allows us to conclude that if $u^*_5>0$ it follows that $u^*_1$ is unbounded,
a contradiction. Hence $u^*_5=0$. If $u^*_1>0$ it follows that $u_2$ is
unbounded, again a contradiction. Hence all the $u_i$ with $1\le i\le 5$
converge pointwise to zero for $t\to\infty$. $\blacksquare$

\section{Stability of steady states}\label{stability}

Consider now the stability of the steady states of the MAd system. The
simplest case to analyse is that of the boundary steady state
$P_0=(0,0,0,0,0,c)$. We first look at the stability of this solution to
homogeneous perturbations, i.e. its stability when considered as a solution of
the MAdh system. The linearization of the MAdh system about an arbitrary
steady state is
\begin{equation}
A=\left[  
  {\begin{array}{cccccc}
     -k_1 & 0 & 0& 0& k_5u_6 & k_5u_5\\
     2k_1 & -k_2u_6-k_6 & 0 & 0 & 0 & -k_2u_2\\
     0    &  k_2u_6 & -k_3 & 0 & 0 & k_2u_2\\
     0 & 0  & k_3 & -k_7-25k_4u_4^4 & 0 & 0\\
     0 & 0 & 0 & 15k_4u_4^4 & -k_5u_6 & -k_5u_5\\
     0 & -k_2u_6 & 0 & 0 & -k_5u_6 & -k_2u_2-k_5u_5-k_8
\end{array}}
\right]
\end{equation}
where we have used the notation $k_8=c$. When this is evaluated at $P_0$ 
it is easily seen that the eigenvalues of the linearization are equal to the
diagonal elements and these are all negative. Hence $P_0$ is a hyperbolic sink
of the MAdh system. To get information about stability to inhomogeneous
perturbations we consider the region $Q_{a,b}$ defined by the inequalities
$0\le L_2\le a$ where $L_2=u_1+\frac12 (u_2+u_3+u_4)+u_5$ and $0\le u_6\le b$.
The aim is to show that for suitable choices of $a$ and $b$ this is an
invariant region in the sense of Chapter 14 of \cite{smoller94}. $L_2$ is a
Lyapunov function considered in \cite{rendall14} and under the evolution of
the MAdh system we have
$\frac{dL_2}{dt}=-\frac16 k_6u_2-\frac12 \left(k_7-k_4u_4^4\right)u_4$.
We need to check that the vector field defining the dynamical system points
into the region $Q_{a,b}$ for suitable choices of $a$ and $b$. In the case of
the part of the boundary defined by $L_2$ it suffices to suppose that
$a\le\frac12\left(\frac{k_7}{k_4}\right)^{\frac14}$. In the case of the part of
the boundary defined by $u_6$ it suffices to require that
$b\le\frac{k_8c}{2(k_2+k_5)a+k_8}$. This means, using Theorem 14.11 of
\cite{smoller94}, that we can cover a neighbourhood of $P_0$ with a
one-parameter family of nested invariant regions which converge to $P_0$ as
the parameter tends to a limiting value. It can be concluded that
$P_0$ is a stable solution of the MAd system with respect to the $L^\infty$
norm. In particular, solutions which start close enough to $P_0$ in the
$L^\infty$ norm are globally bounded. For solutions of this type it can be
proved just as in the last section that $u_i\to 0$ pointwise as $t\to\infty$
for $1\le i\le 5$.

It turns out that all positive steady states of the MAd model are unstable
under general perturbations. This will now be proved following a strategy
used in \cite{marciniak17}. The first step is to prove spectral instability.
The linearization of the system about a steady state takes the form
$v_t={\cal A}v$. Spectral instability means that the spectrum of ${\cal A}$
has a non-empty intersection with the half-plane ${\rm Re}\ z>0$. We think
of $\cal A$ as an unbounded linear operator from $(W^{1,2})^6$ to itself with
domain $(W^{1,2})^5\times W^{3,2}_N$, where $W^{3,2}_N$ is the set of functions
belonging to the Sobolev space $W^{3,2}$ satisfying Neumann boundary
conditions. If spectral instability holds then it follows by Theorem 1 of
\cite{shatah00} that the steady state is (nonlinearly) unstable. Let $A_0$ be
the submatrix of $A$ consisting of the first five rows and columns. Then 
\begin{equation}
{\cal A}=\left[  
{\begin{array}{cc}
A_0 & a\\
b  & {\cal E}
\end{array}}
\right]
\end{equation}
where $a=[k_5u_5,\ -k_2u_2,\ k_2u_2,\ 0,\ -k_5u_5]^T$, 
$b=[0,\ -k_2u_6,\ 0,\ 0,\ -k_5u_6]$ and
${\cal E}=-k_2u_2-k_5u_6-k_8+D_{\rm ATP}\frac{\partial^2}{\partial R^2}$. The
determinant of $A_0$ is positive and
all the coefficients of its characteristic polynomial except the constant term
have the same sign. It follows using Descartes' rule of signs that $A_0$ has
precisely one positive eigenvalue $\lambda_0$. It can be read off from the
form of the matrix that all components of the eigenvector corresponding to
the eigenvalue $\lambda_0$ have the same sign. We can choose them to be
positive. Evaluating the eigenvalue at a given steady
state leads to a positive smooth function $\lambda_0(R)$. Let its infimum
and supremum be denoted by $\lambda_-$ and $\lambda_+$, respectively. The
following is a modification of Theorem 4.5 of \cite{marciniak17}.

\noindent
{\bf Proposition 1} The interval $[\lambda_-,\lambda_+]$ is contained in the
spectrum of ${\cal A}$.

\noindent
{\bf Proof} Let $\lambda\in [\lambda_-,\lambda_+]$ and consider the operator
${\cal A}-\lambda I$. To prove the proposition it suffices to show that this
operator considered as a mapping from $(W^{1,2})^5\times W^{3,2}_N$ to
$(W^{1,2})^6$ does not have a bounded inverse. If it had a bounded inverse then
there would exist a constant $K$ with
\begin{equation}
  \|\bar v\|_{W^{1,2}}+\|v_6\|_{W^{3,2}}
  \le K(\|(A_0-\lambda I)\bar v+v_6 a\|_{W^{1,2}}
  +\|b\cdot \bar v+({\cal E}-\lambda)v_6\|_{W^{1,2}}).
\end{equation}
for all $\bar v\in W^{1,2}$ and $v_6\in W^{3,2}_N$.
There exists a point $R_0$ such $\lambda_0(R_0)=\lambda$ so that
$A_0(R_0)-\lambda I$ has a non-trivial kernel. Let $w(R)$ be a nowhere vanishing
smooth solution of $(A_0(R)-\lambda_0(R))w(R)=0$. Choose $\bar v=\phi w$ for a
smooth function $\phi$. Then $(A_0-\lambda I)\bar v=(\lambda_0-\lambda)\bar v$.
Suppose that the support of $\phi$ is contained in an interval of
length $\epsilon$ and that on that interval $|\lambda_0-\lambda|<\epsilon$.
There is a smooth
function $\psi$ which is identically one on the support of $\bar v$, has
support in an interval $I$ of length $2\epsilon$ and satisfies
$\|\psi\|_{L^\infty}\le 1$ and $\|D\psi\|_{L^\infty}\le\epsilon^{-1}$. Let
$f=(\lambda_0-\lambda)\bar v$. Then
\begin{eqnarray}
&&\|f\|_{L^2}=\|\psi f\|_{L^2}\le\|\psi\|_{L^2}\|f|_I\|_{L^\infty}\le C\epsilon^2,\\
&&\|Df\|_{L^2}=\|D(\psi f)\|_{L^2},\\
&&\|\psi Df\|_{L^2}\le \|\psi\|_{L^2}\|Df|_I\|_{L^\infty}\le C\epsilon,\\
&&\|(D\psi)f\|_{L^2}\le \|D\psi\|_{L^2}\|f|_I\|_{L^\infty}\le C\epsilon^{\frac12}.
\end{eqnarray}
It follows that $\|f\|_{W^{1,2}}\le C\epsilon^{\frac12}$. Let $v_6$
be any smooth function with support in $I$. Since $w$ is
positive it follows that $b\cdot w$ is negative and we can choose
$\phi=-\frac{({\cal E}-\lambda)v_6}{b\cdot w}$. It follows that
\begin{equation}
(1-C\epsilon^{\frac12})\|\bar v\|_{W^{1,2}}+\|v_6\|_{W^{3,2}}\le C\|v_6\|_{W^{1,2}}.
\end{equation}
Since $v_6$ was arbitrary this gives a contradiction for $\epsilon$
sufficiently small. $\blacksquare$ 

\noindent
{\bf Theorem 4} All smooth positive steady states of the MAd system are
nonlinearly unstable in the Sobolev space $W^{1,2}$.

\noindent
{\bf Proof} The full nonlinear system can be written in the form
$v_t={\cal A}v+F(v)$ where $F$ is at least quadratic in $v$. More precisely,
$F(v)$ is a sum of terms, each of which is of one of the following types. The
first type is a smooth function of $\bar v$. Since the space dimension is one
$W^{1,2}$ is continuously embedded in $L^\infty$. It follows from the Moser
estimates \cite{taylor96} that for all $\bar v$ in a fixed ball in
$(W^{1,2})^5$ this type of term can be bounded by $C\|\bar v\|^2_{W^{1,2}}$.  
for a constant $C>0$. The second type is proportional to an expression of the
form $v_iv_6$ with $1\le i\le 5$. It can be bounded by
$C\|\bar v\|_{W^{1,2}}\|v_6\|_{L^2}$. Thus
if $\cal A$ is considered as an operator on $(W^{1,2})^6$ with domain
$(W^{1,2})^5\times W^{3,2}_N$ it satisfies the hypothesis (iii) of Theorem 1 of
\cite{shatah00}. In addition this operator generates a strongly continuous
semigroup on this space. To see this note that it is the sum of a bounded
operator with the operator $(0,\Delta)$. Thus it suffices to know that the
Laplacian generates a strongly continuous semigroup on $W^{1,2}$ with domain
$W^{3,2}_N$. The statement of the theorem follows. $\blacksquare$

It should that recently a much more general theorem of this type has been
proved in \cite{cygan21}. Note also that an instability result for
steady states of the MAd system with periodic boundary conditions can be
proved in the same way.

\section{Oscillations}\label{oscillations}

\begin{figure}[htbp]
	\centering
	\includegraphics[width=0.7\textwidth]{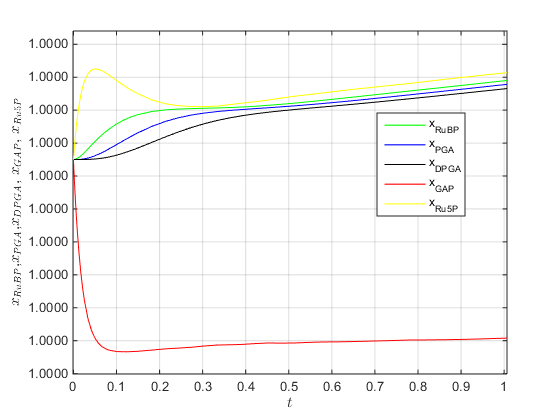}
	\caption{Non-monotone behaviour near $P_1$\label{fig:osc1}.}
\end{figure}

\begin{figure}[htbp]
	\centering
	\includegraphics[width=0.7\textwidth]{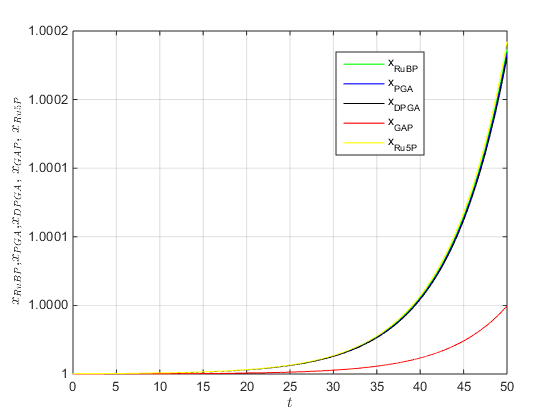}
	\caption{Long-time behaviour of the solution in Fig. \ref{fig:osc1}
          \label{fig:fig2}.}
\end{figure}

In this section we address the question of the occurence of oscillations in
the MA and MAdh systems. Consider a steady state of the MA system. If the
linearization $A$ of the system about that point has eigenvalues with non-zero
imaginary parts then this indicates the presence of oscillations. In general we
know that $A$ has exactly one positive eigenvalue. An example will now be
presented where there are non-real eigenvalues. To make the calculations as
simple as possible we assume that there is a steady state where all
concentrations $u_i$, $1\le i\le 5$, are equal to one. The rate constants
$k_1$ and $k_2$ are prescribed so that $\frac53 k_1< k_2<2k_1$. Then it follows
from the equations for steady states that $k_3=k_2$, $k_4=\frac13 k_1$,
$k_5=k_1$, $k_6=2k_1-k_2$ and $k_7=k_2-\frac53 k_1$. Note that all $k_i$
defined in this way are positive. When they are defined in this way
$(1,1,1,1,1)$ is a steady state, which we call $P_1$. To get a concrete
example we choose $k_1=6$ and $k_2=11$. Then $k_3=11$, $k_4=2$, $k_5=6$,
$k_6=1$ and $k_7=1$. The linearization is
\begin{equation}
  \left[
{\begin{array}{ccccc}    
  -6 & 0 & 0 & 0 & 6\\
12 &-12& 0 & 0 & 0\\
0  &11 &-11& 0 & 0\\
0  & 0 & 11&-51& 0\\  
0  & 0 & 0 & 30 &-6
\end{array}}
\right]
\end{equation}
It seems difficult to obtain information about the eigenvalues of this matrix
by hand but a computer calculation shows that there are two negative real
eigenvalues and two complex eigenvalues with negative real parts. Thus this
is an example where it can be expected that there are solutions which are
oscillatory in the sense that some of the concentrations not monotone. Note for
comparison that in a simpler analogue of this model due to Hahn \cite{hahn91}
the eigenvalues of the linearization about a steady state are always real
\cite{obeid19}. Fig. \ref{fig:osc1} shows an example where the concentrations
are not monotone.
As shown in Fig. \ref{fig:fig2} these solutions are such that all
concentrations later become monotone and much larger. Presumably the
solution approaches the one-dimensional unstable manifold of $P_1$.

\begin{figure}[htbp]
	\centering
	\includegraphics[width=0.7\textwidth]{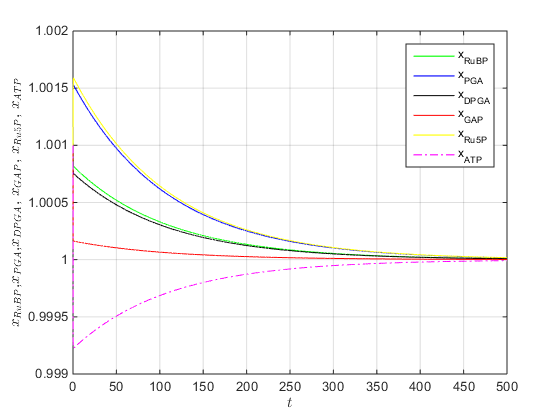}
	\caption{Long-time behaviour of a solution approaching the point $P_2$
          \label{fig:fig3}.}
\end{figure}

A similar method can be applied to the MAdh model. Suppose that at a steady
state all concentrations are equal to one and that $c=2$. The reaction
constants can be chosen as $(6,11,11,2,6,1,1,17)$. We call this steady state
$P_2$. In this case the linearization is
\begin{equation}
  \left[
{\begin{array}{cccccc}    
  -6 & 0 & 0 & 0 & 6& 6\\
12 &-12& 0 & 0 & 0& -11\\
0  &11 &-11& 0 & 0& 11\\
0  & 0 & 11&-51& 0& 0\\
0  & 0 & 0 & 30 &-6&-6\\
0  &-11& 0 & 0 &-6&-34   
\end{array}}
\right]
\end{equation}
Computer calculations show that this matrix has four negative eigenvalues and
two non-real eigenvalues with negative real part. Thus it is a hyperbolic sink.
In a simulation on a long time-scale (see Fig. \ref{fig:fig3}) the solution
appears
to approach the sink in a monotone manner. Presumably it is approaching $P_2$
along the direction of eigenvalue with smallest real part (which is real) and
the influence of the other eigenvalues is not visible.

According to the
analysis of \cite{rendall14} there should be a second steady state, call it
$P_3$, for these values of the parameters. With these parameters
\begin{eqnarray}
&&x_{\rm GAP}^4=\frac{(11x_{\rm ATP}+1)}{2(11x_{\rm ATP}-5)},\label{P3.1}\\
&&x_{\rm ATP}=2-\frac{16}{17}x_{\rm GAP}^5-\frac{1}{17}x_{\rm GAP}.\label{P3.2}
\end{eqnarray}
Combining (\ref{P3.1}) and (\ref{P3.2}) gives
\begin{equation}
352x_{\rm GAP}^9-154x_{\rm GAP}^5-578x_{\rm GAP}^4-11x_{\rm GAP}+391=0.
\end{equation}
Since $x=1$ is a solution Descartes' rule of signs shows that there is precisely
one other positive root. A numerical calculation shows that it is
approximately $0.98988$. It follows that the coordinates of this steady state
are approximately $(0.95042, 0.91448, 0.95400, 0.98988, 0.90752, 1.04726)$.
We already know that there cannot be more than two steady states and so
$P_2$ and $P_3$ are the only ones in this case.

In simulations we were not able to obtain evidence for the presence of a large
number of oscillations in solutions of the MA oder MAdh systems. To obtain
some more insights into the dynamics it is useful to plot two variables
against each other rather than looking directly at their time dependence.
Fig. 4 gives a plot of the variables $x_{\rm GAP}$ and $x_{\rm Ru5P}$ against each
other for a solution which starts near $P_2$. The curve
obtained exhibits what looks like a corner and it would be interesting to
know how it can be interpreted. One idea is that it could represent a situation
where a solution passes close to a saddle point. However this cannot be the
right explanation since the only steady states which exist, $P_2$ and $P_3$,
are not in the region being plotted. A second alterative is that it
repesents a point where the solution comes close to some kind of slow
manifold. This is consistent with the qualitative description of the
dynamics given above. Understanding whether there are damped oscillations,
periodic solutions or sustained oscillations related to a strange
attractor will require much more extensive numerical investigations than
have been done in the context of the present paper.

\begin{figure}[htbp]
	\centering
	\includegraphics[width=0.7\textwidth]{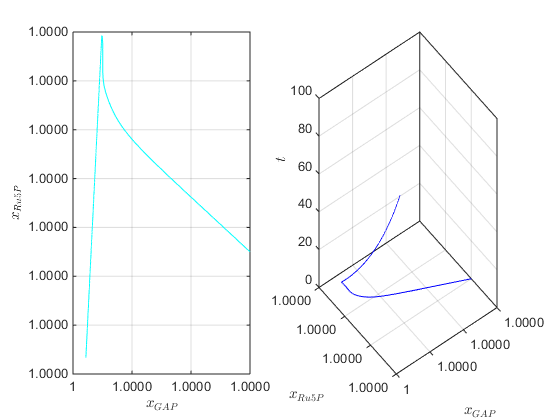}
	\caption{Plot in the $x_{\rm GAP}$-$x_{\rm Ru5P}$ plane of a solution starting near $P_2$
          \label{fig:fig4}.}
\end{figure}

We now compare the calculations of this section with some statements in
\cite{grimbs11}. In Table 5 of that reference
the concentrations in a steady state of the MAdh
system are given. The rate constants for which this is supposed to be a
steady state are not given. It is, however, possible, to get some
information about those rate constants. Note that
\begin{eqnarray}
&&0<k_7x_{\rm GAP}=k_2x_{\rm PGA}x_{\rm ATP}-\frac35k_1x_{\rm RuBP}\\
&&0<k_6x_{\rm PGA}=2k_1x_{\rm RuBP}-k_2x_{\rm PGA}x_{\rm ATP}
\end{eqnarray}
Hence if $\alpha=\frac{x_{\rm RuBP}}{x_{\rm PGA}x_{\rm ATP}}$ then
$\frac53\alpha k_1<k_2<2\alpha k_1$. For the concentrations given in
\cite{grimbs11} the constant $\alpha$ is approximately equal to $1.43$.

\section{Conclusions and outlook}

In this paper we have investigated some aspects of the dynamics of solutions
of a system of reaction-diffusion equations (the MAd system) modelling the
Calvin cycle of photosynthesis which takes the diffusion of ATP into
account. We also compared this system with a related system of ODE (the MA
system) where ATP is not allowed to diffuse. It had been suggested that
the existence of more than one steady state of the MAd system could help to
explain the observation of more than one steady state in experiments. We
proved that for suitable values of the parameters this system does admit
infinitely many inhomogeneous steady states. At the same time their
biological relevance is limited by the fact that we proved that all
positive steady states of the MAd system are nonlinearly unstable. In
fact this is a frequent feature of reaction-diffusion systems where not all
species diffuse. It is a phenomenon which may not be detected by the study
of eigenvalues of the linearization about the steady state since the
instability we exhibit is generated by the continuous spectrum of the
linearized operator.

There are a number of mathematical questions about the global behaviour of
solutions of the MAd and MA models which remain open, since we were only
able to obtain limited results on them. Some of these questions will now be
listed. Are all solutions of the MAd system bounded in $L^1$? If so, are they
all bounded in $L^\infty$? Do there exist periodic solutions of the MA system
or periodic spatially homogeneous solutions of the MAd system? Do there exist
chaotic solutions of the MA system or chaotic spatially homogeneous solutions
of the MAd system. Answers to these questions could contribute to the
general task of understanding the long-time behaviour of general solutions of
the MAd and MA systems. They might also help to give an alternative solution
to the biological question which motivated this research. The original
idea was that an experimentally observed steady state might be spatially
inhomogeneous and that this might not be evident since the quantities measured
are spatial averages over a certain region. A variant of this is that the
'steady state' might be a persistent oscillation which is not recognized as
such because the quantities measured are temporal averages over certain time
intervals.

\vskip 10pt\noindent
{\it Acknowledgements} We thank Anna Marciniak-Czochra and Patrick Tolksdorf
for helpful discussions.


\end{document}